\documentclass[11pt,tightenlines,eqsecnum,floats,aps,
amsmath,amssymb,nofootinbib,prd,shownopacs,floatfix,longbibliography]{revtex4-1}
\usepackage{graphicx}
\usepackage{epstopdf}
\usepackage{latexsym}
\usepackage{amssymb}
\usepackage{amsmath}
\usepackage{color}
\usepackage{mathrsfs}
\usepackage{xparse}
\usepackage{float}
\usepackage{mathtools}
\usepackage{multirow}
\usepackage[center]{subfigure}

\usepackage{xcolor}

\graphicspath{{figs/}} 

\begin{document}
\renewcommand\arraystretch{2}
 \newcommand{\bq}{\begin{equation}}
 \newcommand{\eq}{\end{equation}}
 \newcommand{\bqn}{\begin{eqnarray}}
 \newcommand{\eqn}{\end{eqnarray}}
 \newcommand{\nb}{\nonumber}
 \newcommand{\cb}{\color{blue}}
    \newcommand{\cc}{\color{cyan}}
     \newcommand{\lb}{\label}
        \newcommand{\cm}{\color{magenta}}
\newcommand{\rc}{\rho^{\scriptscriptstyle{\mathrm{I}}}_c}
\newcommand{\rd}{\rho^{\scriptscriptstyle{\mathrm{II}}}_c}
\NewDocumentCommand{\evalat}{sO{\big}mm}{%
  \IfBooleanTF{#1}
   {\mleft. #3 \mright|_{#4}}
   {#3#2|_{#4}}%
}

\newcommand{\PRL}{Phys. Rev. Lett.}
\newcommand{\PL}{Phys. Lett.}
\newcommand{\PR}{Phys. Rev.}
\newcommand{\CQG}{Class. Quantum Grav.}
\newcommand{\parallelsum}{\mathbin{\!/\mkern-5mu/\!}}

\title{The primordial angular power spectrum from the alternative mass function in loop quantum cosmology}

\author{Abolhassan Mohammadi}
\email{abolhassanm@zjut.edu.cn}
\affiliation{Institute for Theoretical Physics \& Cosmology, Zhejiang University of Technology, Hangzhou, 310023, China}
\affiliation{United Center for Gravitational Wave Physics, Zhejiang University of Technology, Hangzhou, 310023, China}

\author{Bao-Fei Li}
\email{Corresponding author: libaofei@zjut.edu.cn}
\affiliation{Institute for Theoretical Physics \& Cosmology, Zhejiang University of Technology, Hangzhou, 310023, China}
\affiliation{United Center for Gravitational Wave Physics, Zhejiang University of Technology, Hangzhou, 310023, China}

\author{Tao Zhu}
\email{zhut05@zjut.edu.cn}
\affiliation{Institute for Theoretical Physics \& Cosmology, Zhejiang University of Technology, Hangzhou, 310023, China}
\affiliation{United Center for Gravitational Wave Physics, Zhejiang University of Technology, Hangzhou, 310023, China}
 
\begin{abstract}

We investigate the cosmological impacts of the alternative effective mass function of the modified Mukhanov-Sasaki equation in loop quantum cosmology, which is obtained from the polymerization of the classical mass function derived in the comoving gauge. This alternative effective mass function is distinct from those in the dressed metric and the hybrid approaches and is able to generate a new structure in the primordial power spectrum. After taking the Starobinsky potential and employing a particular polymerization ansatz for the inverse Hubble rate, the effective mass function is characterized by a free parameter $\xi$. When $\xi \ge 0.1$, there appears a wave-packet structure in the region preceding the almost scale invariant regime of the power spectrum and both the location and the height of the wave packet are affected by the choice of $\xi$. For the angular power spectrum, we find $\xi=0.2$ provides the best-fit curve to the result from the $\Lambda$CDM model. Our study presents a concrete example in which the fine structure of the primordial power spectrum sensitively relies on the parameters in the polymerization ansatz. 

\end{abstract}
\maketitle
\section{Introduction}
\lb{sec:intro}

One of the most popular paradigms for describing the early universe is the inflationary scenario, which not only accounts for many long-standing problems of the standard Big Bang model but also predicts the cosmological perturbations that serve as seeds for the formation of the large-scale structure of the universe. However, the classical inflationary spacetime based on general relativity is still past-incomplete \cite{Borde:2001nh} as it would inevitably encounter the big bang singularity at which physical observables, such as the spacetime curvature, the energy density, etc, become infinite. The existence of the singularity is a generic feature of the spacetime \cite{hawking2023large}, which is linked with the limitations of the classical theory beyond which quantum cosmological theories incorporating either the quantum gravity effects or the exotic matter content are expected to play the essential role. One of the candidate theories for quantum cosmology in which the big bang singularity is generically resolved is loop quantum cosmology (LQC) \cite{Ashtekar:2011ni,Li:2023dwy} which applies techniques of loop quantum gravity (LQG) \cite{Thiemann:2007pyv} to numerous cosmological spacetimes, resulting in a generic resolution of all the strong singularities due to the quantum geometric effects \cite{Ashtekar:2006wn,Ashtekar:2007em,Singh:2009mz,Singh:2011gp,Singh:2014fsy,Saini:2017ggt,Saini:2017ipg}. 

In LQC, when the energy density reaches the critical value in the Planck regime, a quantum bounce takes place and connects a contracting branch with an expanding branch. Although the quantum theory of LQC is formulated in terms of the quantum Hamiltonian operator which leads to a quantum difference equation \cite{Ashtekar:2006wn}, numerical simulations have shown that the dynamics generated by the quantum difference equation can be well approximated by the effective dynamics which is governed by an effective Hamiltonian constraint \cite{Taveras:2008ke,Diener:2014mia, Diener:2014hba, Diener:2017lde, Singh:2018rwa}. Using the effective dynamics, many different aspects of the inflationary scenario as well as other alternative scenario, such as the cyclic universe scenario, have been extensively studied, focusing on the extension of these scenarios into the Planckian regime and the unique patterns of the evolution of the background universe in each scenario \cite{Singh:2006im,Mielczarek:2010bh,Ashtekar:2011rm,Linsefors:2013cd,Corichi:2013kua,Bonga:2015kaa,Zhu:2017jew,Li:2020pww,Giesel:2020raf,Motaharfar:2021gwi, Sharma:2019okc, Wu:2018sbr, Jin:2018wdx, Zhu:2015owa, Zhu:2015xsa, Zhu:2015ata, Zhu:2016dkn, Zhu:2017onp, Li:2018vzr}.  

On the other hand, to investigate the potential observable signals arising from the evolution of the cosmological perturbations on the quantum background spacetime in LQC, there exist four main approaches, namely the dressed metric approach \cite{Agullo:2012sh,Agullo:2012fc,Agullo:2013ai}, the hybrid approach \cite{Fernandez-Mendez:2012poe,Fernandez-Mendez:2013jqa,Gomar:2014faa,Gomar:2015oea,Martinez:2016hmn,ElizagaNavascues:2020uyf}, the deformed algebra approach \cite{Bojowald:2008gz,Cailleteau:2012fy,Cailleteau:2011kr} and the separate universe approach \cite{Wilson-Ewing:2015sfx}. The first two approaches are built on a similar line of construction, with the background degrees of freedom being loop quantized and the linear perturbations Fock quantized. As a result, the modified Mukhanov-Sasaki equations in these two approaches differ by different forms of the effective mass function \cite{Iteanu:2022zha}. Moreover, it can be shown that the effective mass functions in two approaches can be obtained from the polymerization of the different forms of the classical mass function which are equivalent on the classical dynamical trajectories \cite{Li:2022evi}. Therefore, different effective mass functions can be understood as arising from the non-commutativity between the polymerization procedures and the evaluation of the Poisson bracket. Recently, in another work \cite{Li:2023res}, an alternative effective mass function has been proposed based on the polymerization of the classical mass function which explicitly depends on the inverse Hubble rate and its derivatives. This classical mass function is originally obtained by using the comoving gauge in the Lagrangian formulation of the linear perturbation theory. The resulting effective mass function differs from those in the dressed metric and the hybrid approaches and depends explicitly on an undetermined function $g$ which relies on the energy density and the polymerization ansatz of the inverse Hubble rate. Since how to extract the cosmological sector from LQG is still an open question, the authors in \cite{Li:2023res} expand the polymerization function $g$ in terms of the powers of the energy density and truncate the series to the first order which depends on an arbitrary parameter $\xi$. This leads to a concrete form of the effective mass function but no continued work on the predictions due to this new mass function has ever been reported.

In this manuscript,  we would like to numerically compute the primordial power spectrum and the angular power spectrum resulting from the new effective mass function proposed in \cite{Li:2023res} and then compare the results with those in the dressed metric and the hybrid approaches to investigate whether the polymerization ansatz of the classical mass function can lead to distinct impacts on the observations. To achieve this purpose, we employ the Starobinsky potential as the inflationary potential and set the proper initial conditions for the background as well as the linear perturbations so that the number of inflationary e-foldings is 68 and the initial state is chosen to be the Bunch-Davies (BD) vacuum in the contracting branch. Although the resulting power spectrum also develops three characteristic regimes, namely the infrared regime, the intermediate oscillatory regime and the almost scale-invariant regime, as in the dressed metric and the hybrid approaches, there also arises a unique feature in the sector of the power spectrum preceding the almost scale invariant regime where a wave-packet structure starts to form and then develops into an apparent property of the power spectrum as the parameter $\xi$ reaches about $0.1$ and increases further. For $\xi\lesssim0.1$, the wave-packet structure disappears.  Moreover, the location and the height of the wave packet also change with the magnitude of $\xi$. On the other hand, with regard to the angular power spectrum, we also find the optimal value of $\xi$ which is $\xi=0.2$ that can lead to the best-fit curve to the result from the $\Lambda$CDM model. While other choices of $\xi$, including $\xi<0$, can lead to obvious deviations from the $\Lambda$CDM model for multipoles $l<10$ and thus are disfavored. Our studies on the primordial power spectrum and the angular power spectrum from the alternative effective mass function in LQC show explicitly that polymerization parameters can lead to substantially discernible signatures on the observations and also conversely observations can in return efficiently help restrict the parameter space of the viable parameters in the polymerization ansatz. 

The manuscript is organized as follows. In Sec. \ref{sec:review}, we review the effective dynamics of the background evolution of LQC as well as the modified Mukhanov-Sasaki equation governing the evolution of the linear cosmological perturbations in LQC. In Sec. \ref{sec:power_spectrum}, we numerically compute the primordial power spectrum and the angular power spectrum resulting from one of the alternative mass functions proposed recently. In Sec. \ref{sec:conclusion}, we summarize our main results.  In the following, we use the Planck units $\hbar=c=1$ while keeping the Newton's constant $G$ explicit.

\section{Review of the modified Mukhanov-Sasaki equation with alternative mass function in loop quantum cosmology} 
\lb{sec:review}

In this section, we briefly review the effective dynamics of the quantum evolution of the background spacetime and the cosmological perturbations in LQC for a spatially flat FLRW universe filled with an inflaton field. Due to the homogeneity and isotropy of the background spacetime, the phase space of this dynamical system is spanned by four degrees of freedom which can be chosen as $\{c,p,\phi,p_\phi\}$, here $c$ and $p$ are the symmetry reduced variables of the Ashtekar-Barbero connection $A_a^i$ and the densitized triad $E^a_i$ which satisfy the fundamental Poisson bracket $\{c,p\}=8\pi G \gamma/3$ with $\gamma$ denoting the Barbero-Immirzi parameter.  The matter sector is composed of an inflaton field $\phi$ and its conjugate momentum $p_\phi$ with the standard Poisson bracket $\{\phi,p_\phi\}=1$. In the $\bar \mu$ scheme of LQC, instead of $c$ and $p$, the geometric sector for a spatially flat FLRW universe is formulated in terms of the physical volume $v$ and its conjugate momentum $b$ which are given respectively by $v=|p|^{3/2}$ and $b=c/\sqrt{|p|}$ \cite{Ashtekar:2006wn}. The momentum $b$ is proportional to the Hubble rate in the classical regime. Moreover, it has been shown by numerical simulations that the evolution of the quantum dynamics in LQC can be faithfully captured by the effective dynamics \cite{Diener:2014mia,Diener:2014hba} governed by an effective Hamiltonian constraint that takes the form
\bq
\lb{Hamiltonian}
\mathcal{H} = -\frac{3v}{8 \pi G \lambda^2 \gamma^2} 
\sin^2\left(\lambda b \right) + \frac{p_\phi^2} { 2 v} + v U(\phi),
\eq 
where $\lambda=2\sqrt{\sqrt{3}\pi \gamma}$ and $U(\phi)$ denotes the potential of the inflaton field. The resulting Hamilton's equations of motion are given by 
\bqn
\dot{v} &=&\frac{3v}{2\lambda \gamma}\sin(2\lambda b), \quad \quad
\dot{b} =- 4 \pi G \gamma \left(\rho+P\right), \label{b_equation} \\ 
\dot{\phi} & = & \frac{p_\phi}{v}, \quad \quad \quad \quad \quad\quad\label{phi_equation}
\dot{p}_{\phi}  = -v U'(\phi), \label{pphi_equation} 
\eqn
here the prime denotes the derivative with respect to the inflaton field, and $\rho$, $P$ are  the energy density and the pressure of the inflaton field given respectively by 
\bq
\rho=\frac{\dot \phi^2}{2} + U(\phi), \quad \quad P = \frac{\dot \phi^2}{2}  - U(\phi).
\eq
To facilitate the comparison with the standard cosmology and extract the physical implications of the effective dynamics, it is also convenient to make use of the modified Friedmann equation in LQC, which takes the form
\bq
H^2 = \frac{8 \pi G}{3}\rho\left( 1 - \frac{\rho}{\rho_\mathrm{c}} \right),
\eq
where $H$ is the Hubble rate, $\rho_\mathrm{c} = 3 / (8\pi G \lambda^2 \gamma^2)$ is the maximum energy density. The big bounce takes place when the energy density reaches the maximum energy density and thus connects a contracting branch to an expanding one. With the help of the effective dynamics, various aspects of the physical implications of the quantum nature of the big bounce have been investigated in great detail (see \cite{Li:2023dwy} for a recent review).   

In addition to the background dynamics, the physical implications of the cosmological perturbations in LQC have been investigated extensively in mainly four different approaches which are the dressed metric approach \cite{Agullo:2012sh,Agullo:2012fc,Agullo:2013ai}, the hybrid approach \cite{Fernandez-Mendez:2012poe,Fernandez-Mendez:2013jqa,Gomar:2014faa,Gomar:2015oea,Martinez:2016hmn,ElizagaNavascues:2020uyf}, the deformed algebra approach \cite{Bojowald:2008gz,Cailleteau:2012fy,Cailleteau:2011kr} and the separate universe approach \cite{Wilson-Ewing:2015sfx}. In the numerical computations of the primordial power spectrum, the first two approaches are frequently employed. Their modified Mukhanov-Sasaki equations for the cosmological linear perturbations can be effectively obtained from the polymerization of the classical mass function. It has been shown that polymerization of the different forms of the classical mass function, which are originally equivalent on the classical trajectories, can result in distinct effective mass functions in the bounce regime that are employed in the dressed metric and the hybrid approaches \cite{Li:2022evi}. Moreover, in one of the previous works \cite{Li:2023res}, a series of new forms of the effective mass function have been proposed based on the classical mass function obtained from using the comoving gauge in the Lagrangian formulation of the cosmological linear perturbation theory which leads to the classical Mukhanov-Sasaki equation in the form
\bq
\lb{classical_MS}
\nu^{\prime \prime}_{ k}+\left(k^2-\frac{z^{\prime \prime}_s}{z_s}\right)\nu_{ k}=0,
\eq
where the prime denotes the derivative with respect to the conformal time and the rescaled Mukhanov-Sasaki variable $\nu_k$ is related with the comoving curvature perturbation $\mathcal R_k$ via $\nu_{ k}=z_s\mathcal R_k$ and $z_s=a\dot \phi/H$. Thus, in this approach, the classical mass function takes the form 
\bq
\lb{classical_mass}
m^2_\mathrm{CG}=-\frac{z^{\prime \prime}_s}{z_s},
\eq
here the index `CG' implies that the above classical mass function is obtained in the comoving gauge. As discussed in \cite{Li:2023res}, in order to obtain the corresponding effective mass function, one can polymerize the inverse Hubble rate in the expression of $z_s$ by employing a general ansatz
\bq
z_s\rightarrow \frac{a\dot {\phi} \sqrt{\rho_c-\rho}}{H}g(\rho),
\eq
here $g$ is a function of the energy density which satisfies the boundary conditions
\bq
\lb{boundary_condition}
\evalat[\Big]{g(\rho)}{\rho=\rho_c}=\mathrm{const}
\quad  \quad \mathrm{and} \quad  \quad  \evalat[\Big]{g(\rho)}{\rho\ll \rho_\mathrm{c}}\rightarrow \frac{1}{\sqrt{\rho_c}}.
\eq
With this polymerization ansatz, the classical mass function (\ref{classical_mass}) can be polymerized into the effective mass function which takes the shape
\bq
\lb{effective_mass}
 m^2_\mathrm{eff}=\Omega^2-\frac{a^{\prime \prime}}{a},
\eq
with the effective potential given by
\bq
\lb{effective_potential}
\Omega^2=a^2\left(U_{, \phi \phi}+(48 \pi G+\delta_a)U+(6H\frac{\dot {{\phi}}}{\rho}+\delta_b)U_{, \phi}+(\delta_c-\frac{48 \pi G}{\rho})U^2+\delta_d \rho^2\right).
\eq
The correction terms $\delta_a$-$\delta_d$ are the additional terms as compared with the effective mass function in the dressed metric approach. In terms of the function $g$, these correction terms assume the form 
\bqn
\delta_a&=&-\frac{24 \pi G \rho}{\rho_c  g }\left( g-2\left(\rho_c-3\rho\right)  g_{,\rho}+8 \rho\left(\rho-\rho_c\right){ g}_{,\rho\rho}\right), \\
\delta_b&=&-12 H \dot{\phi}\frac{ g_{,\rho}}{ g}, \\
\delta_c&=&\frac{24 \pi G}{\rho_c  g }\left( g+2\rho_c  g_{,\rho}+4 \rho\left(\rho-\rho_c\right){g}_{,\rho\rho}\right), \\
\delta_d&=& -\frac{48 \pi G}{\rho_c  g }\left(\left(2\rho_c-3\rho\right)  g_{,\rho}+2 \rho\left(\rho_c-\rho\right){ g}_{,\rho\rho}\right).
\eqn
It should be noted that there exists no special form of the function $g$ that is able to recover the effective mass function in the hybrid or the dressed metric approach. Therefore, the formula (\ref{effective_mass}) gives rise to a series of new effective mass functions whose exact forms depend on the choice of the function $g$. Since the extraction of the cosmological sector from LQG is still an open question, as attempts for a better understanding of the relation between the effective mass function and the shape of the primordial power spectrum, we discuss a number of feasible polymerization ansatz and explore their potential phenomenological impacts on the cosmological observations. To be specific, in the following, we focus on the simplest choice of the function $g$ proposed in \cite{Li:2023res}. In this ansatz, the function $g$ is given by 
\bq
\lb{poly_ansatz}
g(\xi)=\frac{1}{\sqrt{\rho_c}}\left(1+\xi \frac{\rho}{\rho_c}\right),
\eq
here $\xi$ is an arbitrary parameter of the model whose phenomenological impacts will be studied in the next section. With this polymerization ansatz, the correction terms now read
\bqn
\delta_a&=&-\frac{24\pi G\rho}{\rho_c+\xi\rho}\left(1-2\xi+7\xi\frac{\rho}{\rho_c}\right),\quad \delta_b=-12 H \dot{ \phi}\frac{\xi}{\rho_c+\xi\rho}, \nb\\
\delta_c&=&\frac{24\pi G}{\rho_c+\xi\rho}\left(1+2\xi+\xi\frac{\rho}{\rho_c}\right),\quad \delta_d=-\frac{48 \pi G \xi}{\rho_c\left(\rho_c+\xi\rho\right)}\left(2\rho_c-3\rho\right).
\eqn
As long as $\xi$ is not equal to negative unity,  all the correction terms are well-defined at the bounce. When compared with the effective mass functions in the dressed metric and the hybrid approaches, the effective potential acquires the $\delta_d$ term which also contributes even if the bounce is dominated by the kinetic energy of the scalar field. To be specific, for a kinetic-energy-dominated bounce, by ignoring the contribution from the potential energy of the scalar field, the effective mass function in the bounce regime can be approximated by
\begin{equation}
    m^2_\mathrm{eff} \approx \frac{8\pi G}{3}a^2\rho \left(1-4\frac{\rho}{\rho_c}\right) 
           - (48 \pi G a^2) \; \frac{\xi \; \rho^2}{(\rho_c + \xi \rho)} \left(2-3\frac{\rho}{\rho_c}\right)
\end{equation}
and at the bounce point, it becomes
\begin{equation}
    m^2_\mathrm{eff}\Big|_{\rho=\rho_c} \approx 8\pi G a^2\rho_c\left(\frac{5\xi-1}{1+\xi}\right).
\end{equation}
The sign of the effective mass function at the bounce is now determined by the value of the $\xi$ parameter. In the next section, we study the cosmological impacts of this new effective mass function by numerically computing the resulting primordial power spectrum as well as the angular power spectrum.

\section{The primordial (angular) power spectrum from the alternative mass function in loop quantum cosmology} 

\lb{sec:power_spectrum}

In this section, we compute the numerical primordial (angular) power spectrum arising from the alternative mass function (\ref{effective_mass}) and (\ref{effective_potential}) with the polymerization ansatz given in (\ref{poly_ansatz}). For the evolution of the background dynamics,  we take the inflationary potential to be the Starobinsky potential
\bq
    U(\phi) = \frac{3 m^2}{32 \pi G } \; \left( 1 - e^{- \sqrt{\frac{16 \pi G}{3}}\; \phi}  \right)^2,
\eq
where $m$ denotes the mass of the inflaton field.
The initial conditions of the background dynamics are set at the bounce point with $v_B = 1$, $b_B=\pi / 2\lambda$. The momentum $p_\phi$ is obtained from the Hamiltonian constraint as the value of the scalar field is determined. After setting the initial conditions and specifying the mass of the scalar field, the background equations are solved numerically.   
With respect to the linear perturbations, the initial states are set in the contracting phase, where all the interested modes are inside the horizon and can be counted as the Bunch–Davies (BD) vacuum states
\bq
\lb{BD_vacuum}
\nu_{k} = \frac{1}{\sqrt{2 k}} e^{-i  k \eta},
\eq
where $\eta$ stands for the conformal time.
Note that the BD vacuum state is the zeroth-order adiabatic state, and one can work with the higher order adiabatic states as the initial states of the linear perturbations. However, an earlier work \cite{Li:2024xxz} has shown that selecting the higher order adiabatic states as the initial states do not affect the scale-invariant regime of the primordial power spectrum. The effect of the higher-order adiabatic states on the resulting power spectrum would be on the infrared and the oscillatory regimes concerning the modes of the order of $k\le \mathcal{O}(0.01)$. Since the modes in the infrared regime and most of the oscillatory regime are currently outside the observable universe, we are more interested in the scale-invariant sector of the power spectrum and the part of the oscillatory regime which just precedes the scale-invariant regime. Therefore, we will simply set the initial state to be the BD vacuum state in the contracting phase and the primordial power spectrum can be obtained from 
\begin{equation}
\lb{power_spectrum}
P_{{\mathcal{R}}_{k}} = \frac{P_{\nu_{k}}}{z_{s}^2} = \frac{k^3}{2\pi^2} \frac{|\nu_{k}|^2}{z_{s}^2},
\end{equation}
where $z_s=a\dot \phi/H$ and the mode function is evaluated at the end of inflation.

\subsection{The primordial power spectrum}
\lb{power_spectrum}

\begin{figure}
\includegraphics[width=8cm]{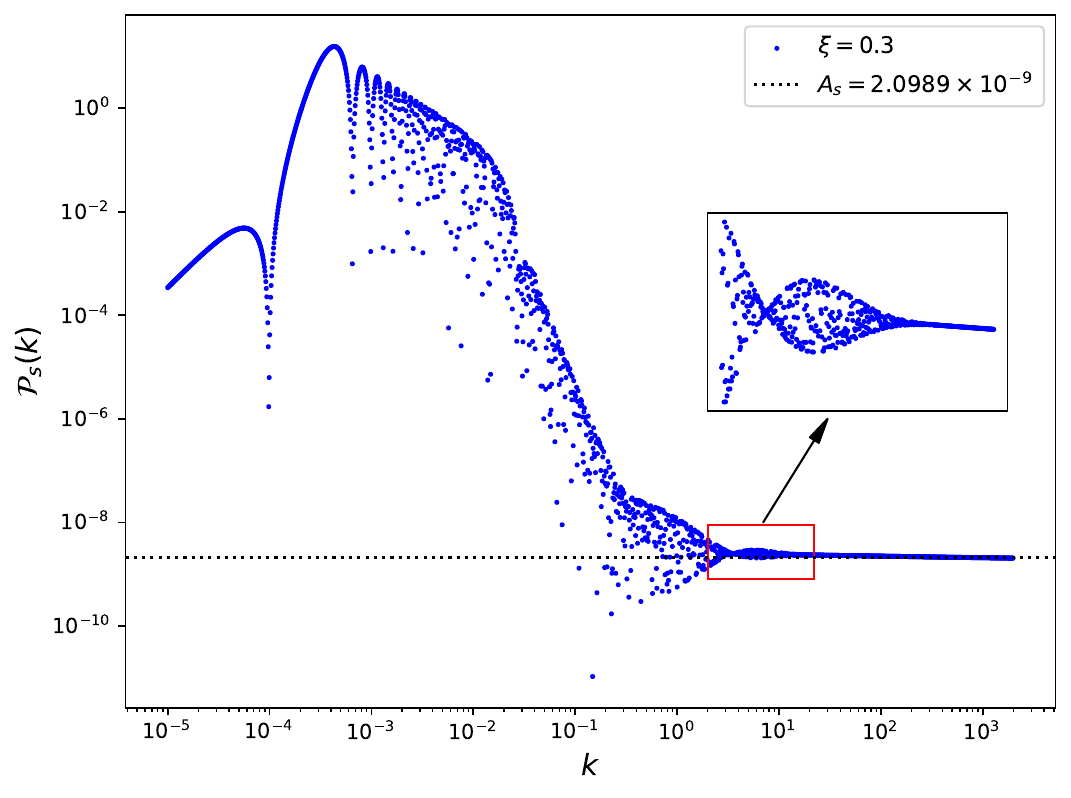}
\caption{A representative example of the primordial power spectrum resulting from the effective mass function (\ref{effective_mass}) and the polymerization ansatz (\ref{poly_ansatz}). To obtain this figure, we set the parameters as $m = 2.455 \times 10^{-6}$,  $\xi = 0.3$ and $\phi_B = -1.427$. The initial time is set at $t_i = -10^6$ in the contracting phase. The inset plot shows the fine structure of the primordial power spectrum in the regime that precedes the almost scale-invariant regime. The black dotted line marks the magnitude of the power spectrum of the pivot mode $k_{\star}/a_0 = 0.05~\mathrm{Mpc^{-1}}$ with $a_0$ being the scale factor at present.}
\label{Ps_vs_k}
\end{figure}

In the actual simulations, we set the initial state at $t_i = -10^6$ and the initial value of the scalar field is chosen to be $\phi_B = -1.427$. The mass of the inflaton is $m = 2.455 \times 10^{-6}$. As a result, the number of the inflationary e-foldings turns out to be $N_\mathrm{inf}=68$. After setting all these parameters, we obtain the representative plot of the resulting power spectrum as given in Fig. \ref{Ps_vs_k}. From this figure, one can find that similar to the primordial power spectrum resulting from the dressed metric and the hybrid approaches, there also exist three distinct regimes which assume the characteristic behavior: the infrared regime with suppressed magnitude which is composed of the scalar modes in the range $k < 10^{-4}$, the amplified oscillatory regime related to the modes in the range $k\in(10^{-4}, 1)$, and the almost scale-invariant regime for the modes $k > 1$. By matching with the amplitude of the power spectrum of the pivot mode observed in CMB, namely $\mathcal{P}_\mathcal{R}(k_\star)=2.0989 \times 10^{-9}$, we find its corresponding comoving wavenumber in the plot turns out to be $k_\star = 706.40$. In general, the value of the comoving wavenumber of the pivot mode depends on the mass of the inflaton field as well as the value of the field at the bounce point, in particular, either choosing a higher $m$ or a smaller $\phi_B$ can lead to a larger $k_\star$. 

%%%==============================
\begin{figure}
\subfigure[\label{ps_k_xi1}]{\includegraphics[width=7.1cm]{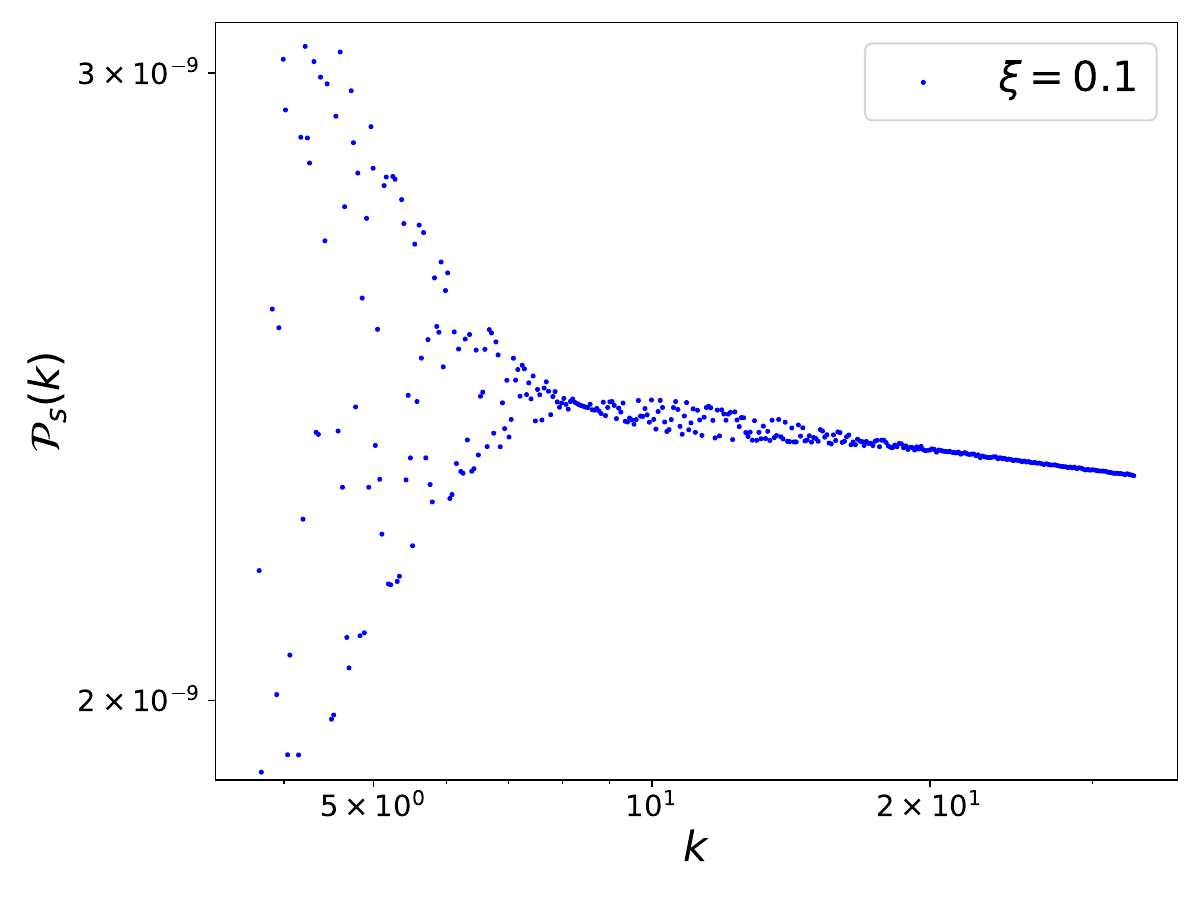}}
\subfigure[\label{ps_k_xi1}]{\includegraphics[width=7.1cm]{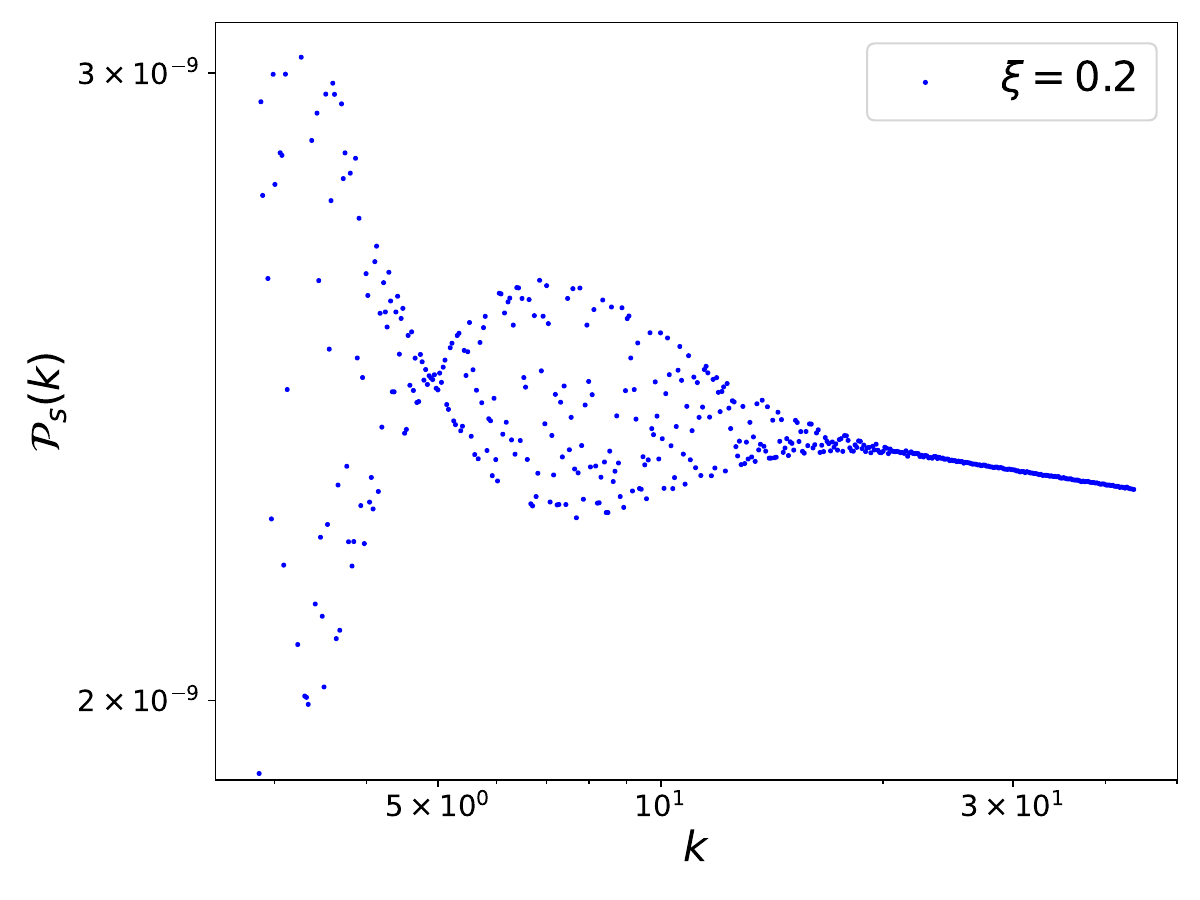}}
\subfigure[\label{ps_k_xi1}]{\includegraphics[width=7cm]{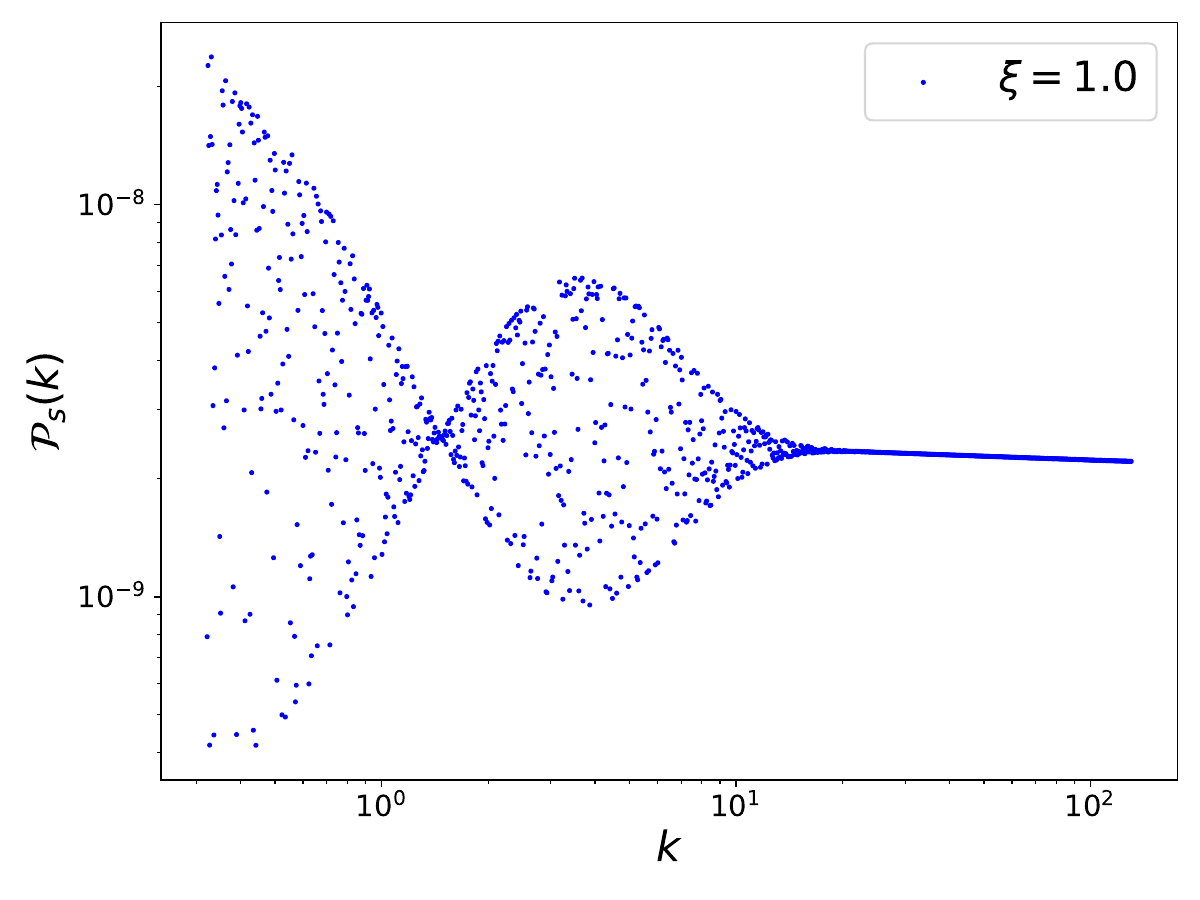}}
\subfigure[\label{ps_k_xi10}]{\includegraphics[width=7cm]{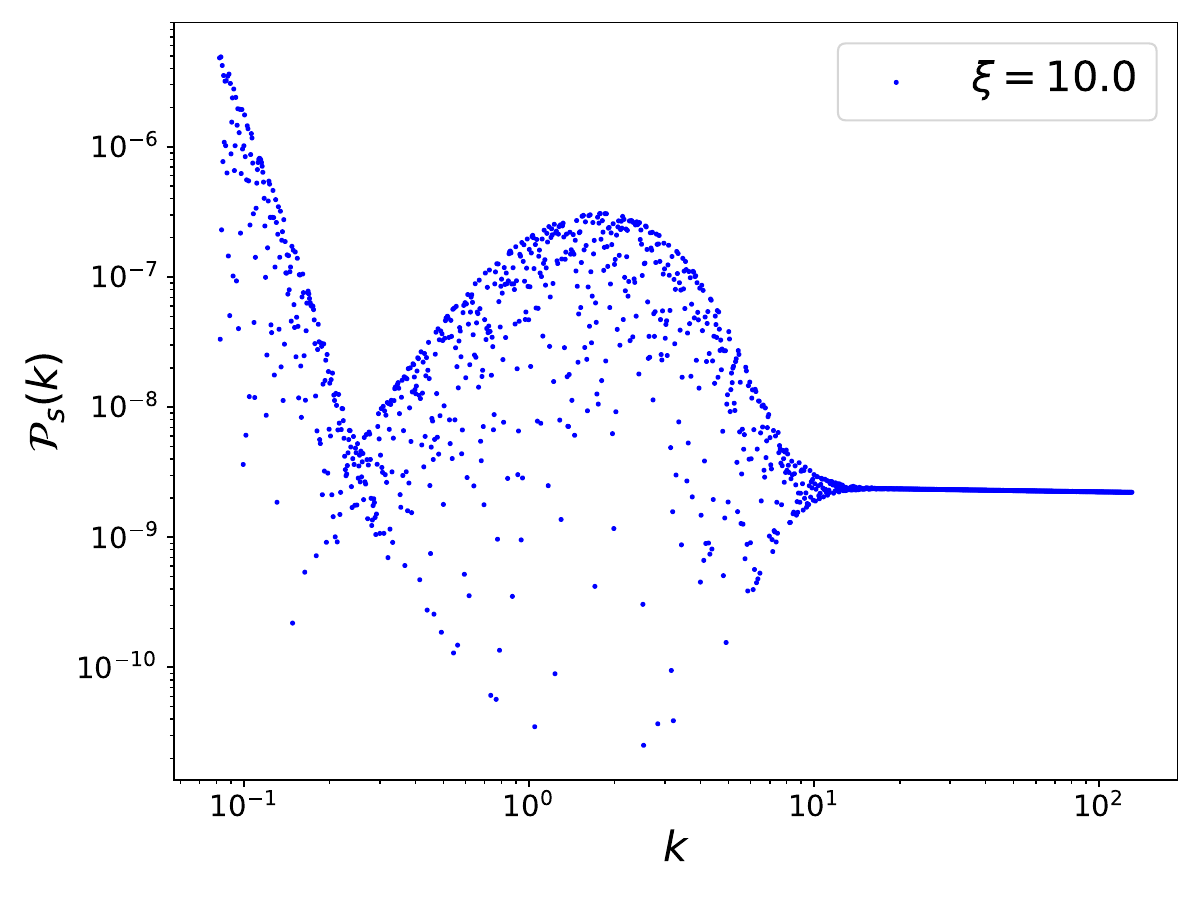}}
\caption{The primordial power spectra resulting from the alternative mass function (\ref{effective_mass}) and the polymerization ansatz (\ref{poly_ansatz}) have been plotted for a) $\xi = 0.1$, b) $\xi = 0.2$, c) $\xi = 1.0$ and d) $\xi = 10.0$. These four subfigures depict the growth of the wave-packet structure in the power spectrum with an increasing $\xi$. All the parameters and the initial conditions are chosen the same as those in Fig. \ref{Ps_vs_k}.  }
\label{ps_k_xi110}
\end{figure}
%%%==================================

\begin{figure}
{\includegraphics[width=8cm]{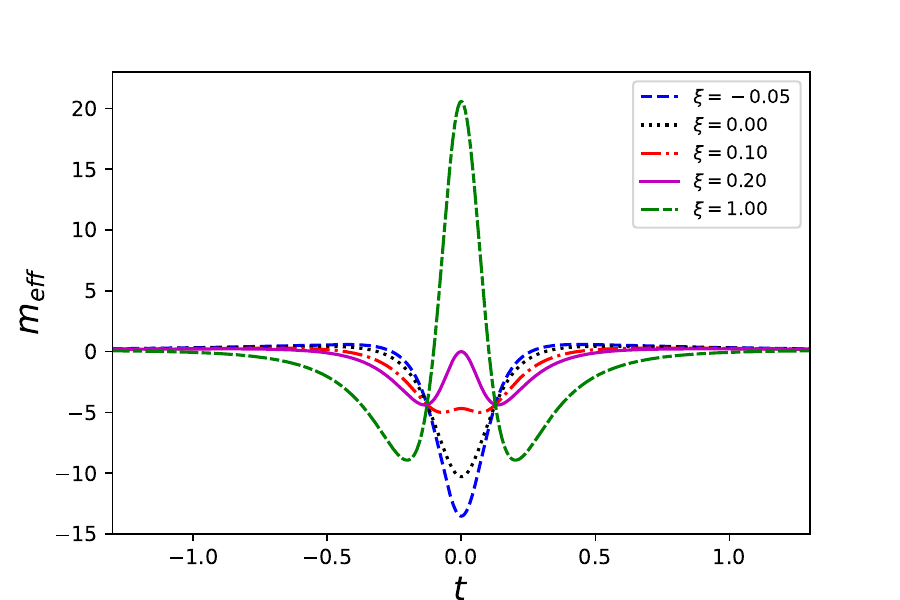}}
\caption{In this figure, we depict the effective mass function near the bounce point where several choices of the parameter $\xi$ can result in the distinct behavior of the effective mass, which impacts the shape of the wave packet in the region of the power spectrum preceding the almost scale invariant regime.}
\label{effective_mass_plot}
\end{figure}

In addition to the familiar pattern of the primordial power spectrum discussed above, in Fig. \ref{Ps_vs_k}, we also observe a new feature that appears in the part of the oscillatory regime that precedes the almost scale-invariant regime. This new feature is also depicted more clearly in the inset plot, where one can find a wave-packet structure connecting the oscillatory regime to the scale-invariant regime. The properties of this wave-packet structure, including the width and the height, solely depend on the magnitude of the parameter $\xi$ and, thus, on the behavior of the effective mass function in the bounce regime. In general, our numerical simulation exhibits that this wave-packet structure does not show up for negative and small positive values of the parameter $\xi$. While for higher values of $\xi$, e.g. $\xi \gtrsim 0.1$, the wave-packet structure appears, and it gets wider and amplified with an increasing $\xi$. To better illustrate the change of the wave packet, we present several more examples in Fig. \ref{ps_k_xi110} to demonstrate the formation and the evolution of the wave-packet structure in the power spectrum with the increasing magnitude of $\xi$. As one can see from the subfigures (a)-(d), the wave-packet structure initially forms in the originally scale-invariant regime, which is right next to the oscillatory regime. As $\xi$ increases, the height of the wave packet becomes larger and its location moves slightly towards the left of the power spectrum. Also, in the subfigures (a)-(c), the dots constituting the wave packet are almost symmetrically distributed with respect to the scale invariant regime, while in subfigure (d), most of the dots now gather in the upper part of the wave packet, which leads to an amplified averaged power spectrum. To understand the relation between the effective mass function and the wave-packet structure in the power spectrum, we plot the effective mass function near the bounce point in Fig. \ref{effective_mass_plot}. As shown in the figure, in the small neighborhood of the bounce point, the effective mass function turns out to be symmetric with respect to the bounce point. When $\xi=-0.05$ and $\xi=0$, the bounce point is a local minimum of the effective mass function. For other cases with $\xi\ge0.10$, the bounce point becomes the local maximum, and there are also two additional local minima in the neighborhood of the bounce point. We tend to believe such a change in the behavior of the mass function with the appearance of multiple extrema points near the bounce potentially leads to the emergence of the wave packet structure observed in Fig. \ref{ps_k_xi110}.

Finally, since the magnitude of $\xi$ would only affect the behavior of the effective mass function in the bounce regime which is known to be responsible for the behavior of the power spectrum only in the oscillatory regime, a change in $\xi$ would not too much affect the scale-invariant regime of the power spectrum. This is consistent with what we have observed in the numerical results. The comoving wavenumber of the pivot mode changes very slightly as $\xi$ parameter changes. The comoving wavenumber of the pivot mode is fixed to be $706.8$ based on the magnitude of the power spectrum. More precisely, for $\xi = -0.1, 0.1, 0.3, 1, 10, 30$, the corresponding values of the comoving wavenumber $k_\star$ are $708.60$, $706.86$, $706.40$, $707.12$, $706.35$, and $705.08$, respectively. In all cases, the deviation from the reference value $k_\star = 706.8$ is less than $0.25\%$, indicating that the position of the pivot scale is practically insensitive to the choice of $\xi$. This confirms that while $\xi$ influences the shape of the power spectrum in the oscillatory regime, it has an ignorable effect on the scale-invariant regime or the determination of the comoving wavenumber of the pivot mode.
 
\subsection{The primordial angular power spectrum}
\lb{angular_power_spectrum}

\begin{figure}
\subfigure[\label{angular_positive}]{\includegraphics[width=8cm]{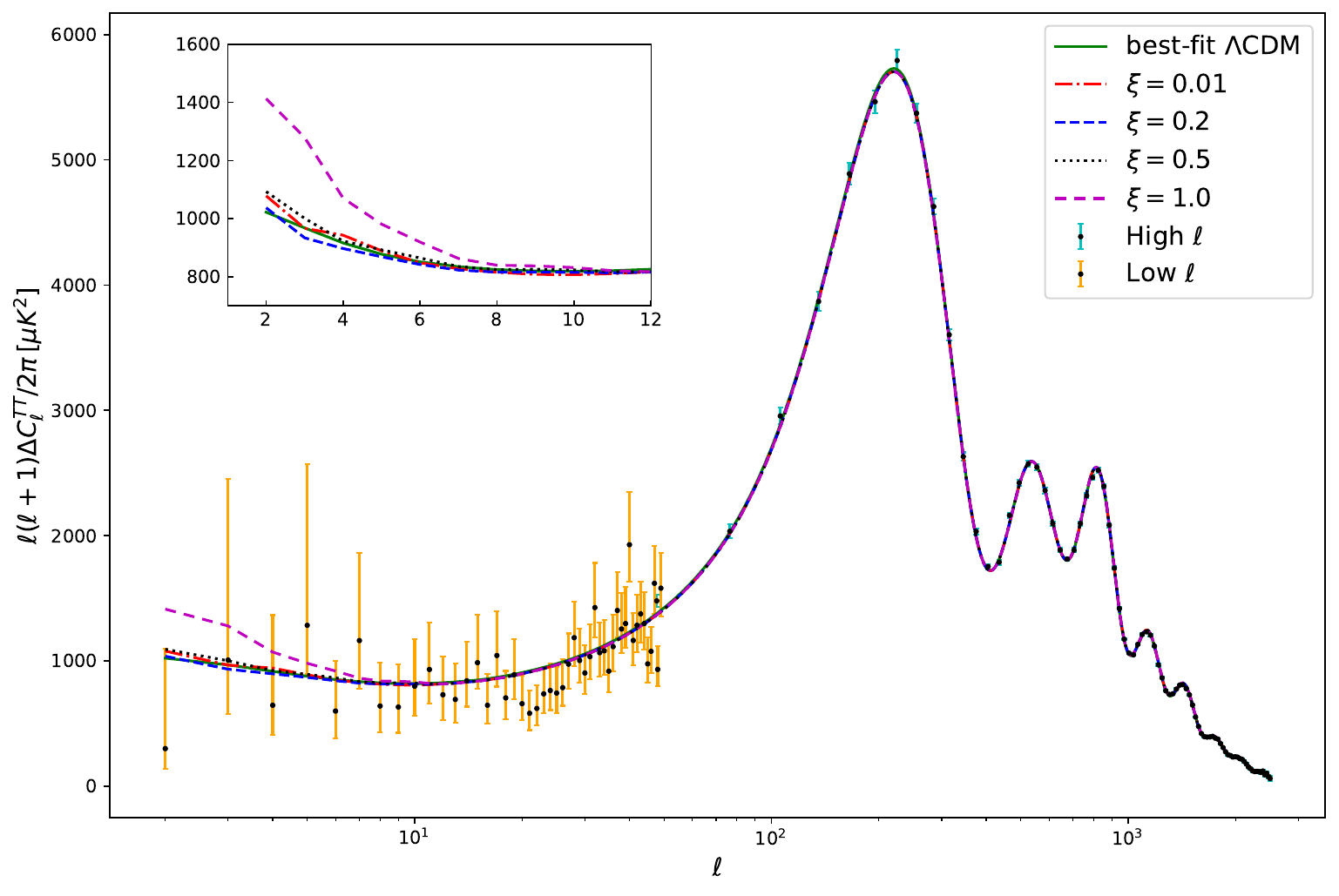}} 
\subfigure[\label{angular_negative}]{\includegraphics[width=8cm]{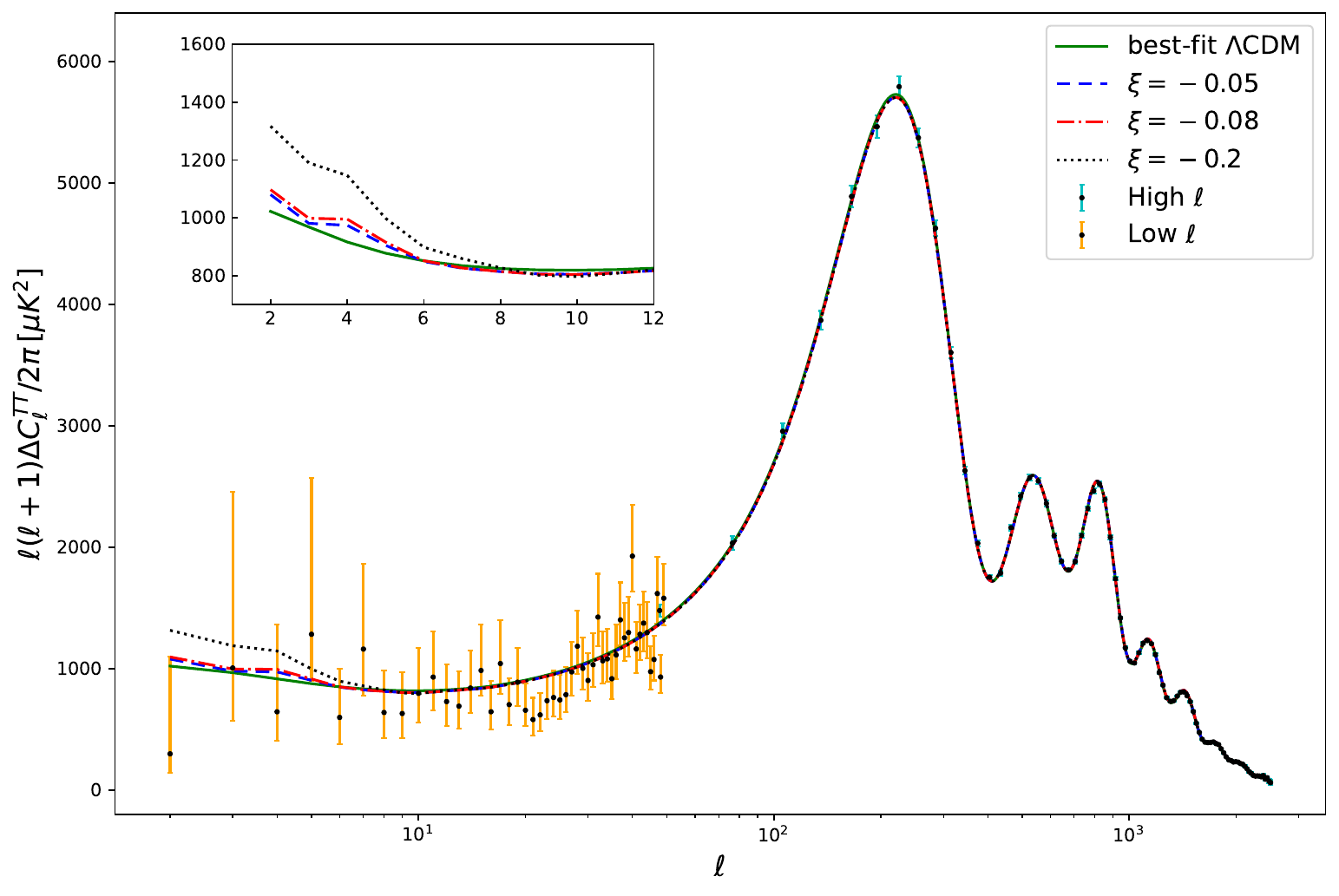}}
\caption{The numerical angular power spectra arising from the effective mass function (\ref{effective_mass}) and the polymerization ansatz (\ref{poly_ansatz}) are presented in the figure. The plot shows the angular power spectra for different values of $\xi$ parameter as: a) positive values of $\xi$, and b) negative values of $\xi$. The black dots show the Planck data for the temperature angular power spectrum with light-blue and orange error bars, respectively, for the high and low multiples $l$. The solid green color represents the angular power spectrum from the best-fit $\Lambda$CDM model. }
\label{angular}
\end{figure}

After obtaining the primordial power spectrum arising from the alternative mass function (\ref{effective_mass}) and the polymerization ansatz (\ref{poly_ansatz}), we now move on to computing the corresponding angular power spectrum by feeding the numerical power spectrum to the CAMB code. The results for the angular power spectrum with different choices of $\xi$ are collectively depicted in Fig.~\ref{angular} in which the Planck-2018 temperature angular power spectrum is represented by a black-dotted curve, where light-blue and orange error bars indicate the uncertainties for high and low multipoles $l$, respectively. The solid green line depicts the best-fit $\Lambda$CDM model to the Planck data (TT, TE, EE, lowE + lensing) \cite{Planck:2018jri,Planck:2018vyg}. In the left panel of the figure, the angular power spectra for positive values of $\xi$, as $\xi = 0.01, 0.2, 0.5, 1.0$, are plotted by the red-dot-dashed, blue-dashed, black-dotted, and purple-dashed lines, respectively. For the multiples larger than $l > 10$, the computed angular power spectrum exhibits excellent agreement with the observational data for all values of $\xi$. However, deviations emerge at smaller multiples. Considering the Planck 2018 data points for low multipoles, the best case is the one that lies below the best-fit $\Lambda$CDM curve, which demonstrates improved consistency with the data. In this regard, the model performs optimally for $\xi = 0.2$. As shown in the inset plot of the left panel of Fig. \ref{angular}, the angular power spectrum for $\xi = 0.2$ remains below the best-fit $\Lambda$CDM for $2<l<9$, but rises and intersects the $\Lambda$CDM curve at $l \approx 2$. A similar behavior is observed for $\xi = 0.01$, where the spectrum lies below the $\Lambda$CDM curve for $l > 6$. However, it crosses the $\Lambda$CDM curve at $l = 6$ and exhibits greater deviation than the $\xi = 0.2$ case. For $\xi = 0.5$, the angular power spectrum initially overlaps with the best-fit $\Lambda$CDM curve but rises around $l = 7$, resulting in larger deviations than those for $\xi = 0.2$ and $0.01$. The deviation grows for $\xi = 1$, where the estimated angular power spectrum begins to diverge from the $\Lambda$CDM model around $l = 10$. 

The theoretical lower limit for the $\xi$ parameter, $\xi > -1$, motivates an investigation of the model for negative values of $\xi$. The right panel of Fig. \ref{angular} illustrates the angular power spectrum for several negative values of $\xi$. To be specific, the blue-dashed, red-dot-dashed, and black-dotted curves represent the angular power spectra corresponding to $\xi = -0.05$, $-0.08$, and $-0.2$, respectively. For multipoles $l > 10$, the resulting angular power spectrum shows excellent agreement with the best-fit $\Lambda$CDM model. However, the spectrum grows for smaller multipoles, leading to deviations from the $\Lambda$CDM curve.

One should also note that there is no linear relationship between the value of $\xi$ and the resulting angular power spectrum. According to our results, the optimal case appears to be $\xi = 0.2$, which exhibits the closest agreement with the observational data, particularly at low multiples. Deviating from this value, either by increasing or decreasing $\xi$, leads to greater deviation from the best-fit $\Lambda$CDM curve. Moreover, our numerical simulation determines that as $\xi$ moves away further from the optimal case $\xi = 0.2$, e.g. $\xi = -0.5$ or $10$,  a significant deviation from the best-fit $\Lambda$CDM curve is produced on small multiples. This non-linear behavior indicates the sensitivity of the model to the $\xi$ parameter, which appears from the polymerization of the inverse Hubble rate in the classical mass function. The obtained result provides some clues about the valid range of the $\xi$ parameter. Although from the theoretical perspective, there is no preferred range for the parameter $\xi$. By comparing to the data, one realizes that only a small range of the $\xi$ parameter is preferred by the observations and the magnitude of the parameter could not be higher than $\mathcal{O}(0.1)$. 

\section{Conclusion}
\lb{sec:conclusion}

In this manuscript, we have investigated the cosmological impacts of the alternative mass function in the modified Mukhanov-Sasaki equation of LQC on the primordial power spectrum and the angular power spectrum. This new effective mass function, initially proposed in \cite{Li:2023res}, is obtained from the polymerization of the classical mass function formulated in the comoving gauge of the Lagrangian formalism. Compared with its counterpart in the dressed metric approach, the new effective mass function acquires four additional correction terms, which arise from the polymerization of the inverse Hubble rate, and their exact forms depend explicitly on the function $g$. Since it is still an open question to extract the cosmological sector from full LQG, we use a tentative polymerization ansatz (\ref{poly_ansatz}) with a free parameter $\xi$ to fix the form of the effective mass function and then investigate its cosmological impacts.  

To obtain the numerical primordial power spectrum, we take the Starobinsky potential as the inflationary potential and choose the specific mass of the inflaton so that the magnitude of the almost scale-invariant regime matches the magnitude of the power spectrum of the pivot mode observed in CMB. The background evolution is set to produce a decent duration of the inflationary phase with its number of e-foldings about 68. Setting the initial states of the cosmological perturbations in the contracting phase, it is then straightforward to obtain the numerical primordial power spectrum.  The result exhibits three regimes of the power spectrum as in the dressed metric and the hybrid approaches, namely the IR regime, the oscillatory regime, and the almost scale-invariant regime. However, a new feature that is distinct from the predictions of the dressed metric and hybrid approaches emerges in the power spectrum. Notably, we have observed a wave-packet structure in the region preceding the scale-invariant regime. Both the location and the height of the wave packet depend on the magnitude of $\xi$, which is able to significantly change the behavior of the effective mass function in the bounce regime. In particular, the wave packet starts to appear in the power spectrum when $\xi\approx0.1$. As $\xi$ increases, the height of the wave packet becomes larger, and it also moves slightly towards the left of the power spectrum. Moreover, the points in the wave packet also begin to cumulate more in the upper part of the wave packet as $\xi$ increases further, say $\xi\ge10$. On the other hand, when $\xi$ decreases, the wave packet structure in the power spectrum disappears, and the power spectrum resembles the one from the dressed metric and the hybrid approaches.   

Regarding the angular power spectrum, we have compared the results from the alternative mass function with the best-fit $\Lambda$CDM model and Planck 2018 data. It is found that our numerical angular power spectrum overlaps the best-fit $\Lambda$CDM for multiples $l > 10$ for all values of $\xi$.  However, deviations arise for small multipoles $l<10$, which depend on the value of $\xi$. The best optimal case we consider is for $\xi = 0.2$, where the angular power spectrum agrees well with the data for $l > 9$ and stays below the best-fit $\Lambda$CDM curve for $l < 9$, and grows and intersects the best-fit $\Lambda$CDM curve at $\l \approx 2$. As the parameter $\xi$ moves away from the optimal value $\xi = 0.2$, we observe a higher deviation of the angular power spectrum from the best-fit $\Lambda$CDM curve for small multiple values. The deviation becomes significant as the parameter $\xi$ gets larger. Our analysis indicates that although no preferred value of $\xi$ is chosen from the theoretical perspective, it should stay within a limited range of around $0.2$ to provide acceptable results about the angular power spectrum. 

\section*{Acknowledgments}
B.-F.~Li is supported by the National Natural Science Foundation of China (NNSFC) with the grant No. 12005186. T.~Zhu is supported by the National Natural Science Foundation of China under Grants No. 12275238, No. 11675143, the National Key Research and Development Program of China under Grant No. 2020YFC2201503, the Zhejiang Provincial Natural Science Foundation of China under Grants No. LR21A050001 and No. LY20A050002, and the Fundamental Research Funds for the Provincial Universities of Zhejiang in China under Grant No. RF-A2019015.


\begin{thebibliography}{10}

\bibitem{Borde:2001nh}
A.~Borde, A.~H. Guth, and A.~Vilenkin,
\newblock {\em Inflationary space-times are incompletein past directions},
\newblock Phys. Rev. Lett. {\bf 90}, 151301 (2003), arXiv:gr-qc/0110012.

\bibitem{hawking2023large}
S.~W. Hawking and G.~F. Ellis,
\newblock {\em The large scale structure of space-time} (Cambridge university
  press, 2023).

\bibitem{Ashtekar:2011ni}
A.~Ashtekar and P.~Singh,
\newblock {\em Loop Quantum Cosmology: A Status Report},
\newblock Class. Quant. Grav. {\bf 28}, 213001 (2011), arXiv:1108.0893.

\bibitem{Li:2023dwy}
B.-F. Li and P.~Singh,
\newblock {\em {Loop Quantum Cosmology: Physics of Singularity Resolution and
  Its Implications}} (2024),  arXiv:2304.05426.

\bibitem{Thiemann:2007pyv}
T.~Thiemann,
\newblock {\em {Modern Canonical Quantum General Relativity, }}Cambridge
  Monographs on Mathematical Physics (Cambridge University Press, 2007).

\bibitem{Ashtekar:2006wn}
A.~Ashtekar, T.~Pawlowski, and P.~Singh,
\newblock {\em Quantum Nature of the Big Bang: Improved dynamics},
\newblock Phys. Rev. D {\bf 74}, 084003 (2006), arXiv:gr-qc/0607039.

\bibitem{Ashtekar:2007em}
A.~Ashtekar, A.~Corichi, and P.~Singh,
\newblock {\em Robustness of key features of loop quantum cosmology},
\newblock Phys. Rev. D {\bf 77}, 024046 (2008), arXiv:0710.3565.

\bibitem{Singh:2009mz}
P.~Singh,
\newblock {\em Are loop quantum cosmos never singular?},
\newblock Class. Quant. Grav. {\bf 26}, 125005 (2009), arXiv:0901.2750.

\bibitem{Singh:2011gp}
P.~Singh,
\newblock {\em Curvature invariants, geodesics, and the strength of
  singularities in Bianchi-I loop quantum cosmology},
\newblock Phys. Rev. D {\bf 85}, 104011 (2012).

\bibitem{Singh:2014fsy}
P.~Singh,
\newblock {\em Loop quantum cosmology and the fate of cosmological
  singularities},
\newblock arXiv preprint arXiv:1509.09182  (2015).

\bibitem{Saini:2017ggt}
S.~Saini and P.~Singh,
\newblock {\em Generic absence of strong singularities in loop quantum
  Bianchi-IX spacetimes},
\newblock Classical and Quantum Gravity {\bf 35}, 065014 (2018).

\bibitem{Saini:2017ipg}
S.~Saini and P.~Singh,
\newblock {\em Resolution of strong singularities and geodesic completeness in
  loop quantum Bianchi-II spacetimes},
\newblock Classical and Quantum Gravity {\bf 34}, 235006 (2017).

\bibitem{Taveras:2008ke}
V.~Taveras,
\newblock {\em Corrections to the Friedmann Equations from LQG for a Universe
  with a Free Scalar Field},
\newblock Phys. Rev. D {\bf 78}, 064072 (2008), arXiv:0807.3325.

\bibitem{Diener:2014mia}
P.~Diener, B.~Gupt, and P.~Singh,
\newblock {\em Numerical simulations of a loop quantum cosmos: robustness of
  the quantum bounce and the validity of effective dynamics},
\newblock Class. Quant. Grav. {\bf 31}, 105015 (2014), arXiv:1402.6613.

\bibitem{Diener:2014hba}
P.~Diener, B.~Gupt, M.~Megevand, and P.~Singh,
\newblock {\em Numerical evolution of squeezed and non-Gaussian states in loop
  quantum cosmology},
\newblock Class. Quant. Grav. {\bf 31}, 165006 (2014), arXiv:1406.1486.

\bibitem{Diener:2017lde}
P.~Diener, A.~Joe, M.~Megevand, and P.~Singh,
\newblock {\em Numerical simulations of loop quantum Bianchi-I spacetimes},
\newblock Class. Quant. Grav. {\bf 34}, 094004 (2017), arXiv:1701.05824.

\bibitem{Singh:2018rwa}
P.~Singh,
\newblock {\em Glimpses of Space-Time Beyond the Singularities Using
  Supercomputers},
\newblock Comput. Sci. Eng. {\bf 20}, 26 (2018), arXiv:1809.01747.

\bibitem{Singh:2006im}
P.~Singh, K.~Vandersloot, and G.~V. Vereshchagin,
\newblock {Non-singular bouncing universes in loop quantum cosmology},
\newblock Phys. Rev. D {\bf 74}, 043510 (2006), arXiv:gr-qc/0606032.

\bibitem{Mielczarek:2010bh}
J.~Mielczarek, T.~Cailleteau, J.~Grain, and A.~Barrau,
\newblock {\em Inflation in loop quantum cosmology: dynamics and spectrum of
  gravitational waves},
\newblock Phys. Rev. D {\bf 81}, 104049 (2010), arXiv:1003.4660.

\bibitem{Ashtekar:2011rm}
A.~Ashtekar and D.~Sloan,
\newblock {\em Probability of Inflation in Loop Quantum Cosmology},
\newblock Gen. Rel. Grav. {\bf 43}, 3619 (2011), arXiv:1103.2475.

\bibitem{Linsefors:2013cd}
L.~Linsefors and A.~Barrau,
\newblock {\em Duration of inflation and conditions at the bounce as a
  prediction of effective isotropic loop quantum cosmology},
\newblock Phys. Rev. D {\bf 87}, 123509 (2013), arXiv:1301.1264.

\bibitem{Corichi:2013kua}
A.~Corichi and D.~Sloan,
\newblock {Inflationary Attractors and their Measures},
\newblock Class. Quant. Grav. {\bf 31}, 062001 (2014), arXiv:1310.6399.

\bibitem{Bonga:2015kaa}
B.~Bonga and B.~Gupt,
\newblock {\em Inflation with the Starobinsky potential in Loop Quantum
  Cosmology},
\newblock Gen. Rel. Grav. {\bf 48}, 71 (2016), arXiv:1510.00680.

\bibitem{Zhu:2017jew}
T.~Zhu, A.~Wang, G.~Cleaver, K.~Kirsten, and Q.~Sheng,
\newblock {\em Pre-inflationary universe in loop quantum cosmology},
\newblock Phys. Rev. D {\bf 96}, 083520 (2017), arXiv:1705.07544.

\bibitem{Li:2020pww}
B.-F. Li, S.~Saini, and P.~Singh,
\newblock {\em Primordial power spectrum from a matter-Ekpyrotic bounce
  scenario in loop quantum cosmology},
\newblock Phys. Rev. D {\bf 103}, 066020 (2021), arXiv:2012.10462.

\bibitem{Giesel:2020raf}
K.~Giesel, B.-F. Li, and P.~Singh,
\newblock {\em Towards a reduced phase space quantization in loop quantum
  cosmology with an inflationary potential},
\newblock Phys. Rev. D {\bf 102}, 126024 (2020), arXiv:2007.06597.

\bibitem{Motaharfar:2021gwi}
M.~Motaharfar and P.~Singh,
\newblock {\em Role of dissipative effects in the quantum gravitational onset
  of warm Starobinsky inflation in a closed universe},
\newblock Phys. Rev. D {\bf 104}, 106006 (2021), arXiv:2102.09578.

\bibitem{Sharma:2019okc}
M.~Sharma, T.~Zhu, and A.~Wang,
\newblock {Background dynamics of pre-inflationary scenario in Brans-Dicke loop
  quantum cosmology},
\newblock Commun. Theor. Phys. {\bf 71}, 1205 (2019), arXiv:1903.07382.

\bibitem{Wu:2018sbr}
Q.~Wu, T.~Zhu, and A.~Wang,
\newblock {\em Nonadiabatic evolution of primordial perturbations and
  non-Gaussinity in hybrid approach of loop quantum cosmology},
\newblock Phys. Rev. D {\bf 98}, 103528 (2018), arXiv:1809.03172.

\bibitem{Jin:2018wdx}
W.-J. Jin, Y.~Ma, and T.~Zhu,
\newblock {Pre-inflationary dynamics of Starobinsky inflation and its
  generization in Loop Quantum Brans-Dicke Cosmology},
\newblock JCAP {\bf 02}, 010 (2019), arXiv:1808.09643.

\bibitem{Zhu:2015owa}
T.~Zhu {\em et~al.},
\newblock {Scalar and tensor perturbations in loop quantum cosmology:
  High-order corrections},
\newblock JCAP {\bf 10}, 052 (2015), arXiv:1508.03239.

\bibitem{Zhu:2015xsa}
T.~Zhu {\em et~al.},
\newblock {Detecting quantum gravitational effects of loop quantum cosmology in
  the early universe?},
\newblock Astrophys. J. Lett. {\bf 807}, L17 (2015), arXiv:1503.06761.

\bibitem{Zhu:2015ata}
T.~Zhu {\em et~al.},
\newblock {Inflationary spectra with inverse-volume corrections in loop quantum
  cosmology and their observational constraints from Planck 2015 data},
\newblock JCAP {\bf 03}, 046 (2016), arXiv:1510.03855.

\bibitem{Zhu:2016dkn}
T.~Zhu, A.~Wang, K.~Kirsten, G.~Cleaver, and Q.~Sheng,
\newblock {Universal features of quantum bounce in loop quantum cosmology},
\newblock Phys. Lett. B {\bf 773}, 196 (2017), arXiv:1607.06329.

\bibitem{Zhu:2017onp}
T.~Zhu, A.~Wang, K.~Kirsten, G.~Cleaver, and Q.~Sheng,
\newblock {\em Primordial non-Gaussianity and power asymmetry with quantum
  gravitational effects in loop quantum cosmology},
\newblock Phys. Rev. D {\bf 97}, 043501 (2018), arXiv:1709.07479.

\bibitem{Li:2018vzr}
B.-F. Li {\em et~al.},
\newblock {\em Preinflationary perturbations from the closed algebra approach
  in loop quantum cosmology},
\newblock Phys. Rev. D {\bf 99}, 103536 (2019), arXiv:1812.11191.

\bibitem{Agullo:2012sh}
I.~Agullo, A.~Ashtekar, and W.~Nelson,
\newblock {\em A Quantum Gravity Extension of the Inflationary Scenario},
\newblock Phys. Rev. Lett. {\bf 109}, 251301 (2012), arXiv:1209.1609.

\bibitem{Agullo:2012fc}
I.~Agullo, A.~Ashtekar, and W.~Nelson,
\newblock {\em Extension of the quantum theory of cosmological perturbations to
  the Planck era},
\newblock Phys. Rev. D {\bf 87}, 043507 (2013), arXiv:1211.1354.

\bibitem{Agullo:2013ai}
I.~Agullo, A.~Ashtekar, and W.~Nelson,
\newblock {\em The pre-inflationary dynamics of loop quantum cosmology:
  Confronting quantum gravity with observations},
\newblock Class. Quant. Grav. {\bf 30}, 085014 (2013), arXiv:1302.0254.

\bibitem{Fernandez-Mendez:2012poe}
M.~Fernandez-Mendez, G.~A. Mena~Marugan, and J.~Olmedo,
\newblock {\em Hybrid quantization of an inflationary universe},
\newblock Phys. Rev. D {\bf 86}, 024003 (2012), arXiv:1205.1917.

\bibitem{Fernandez-Mendez:2013jqa}
M.~Fern\'andez-M\'endez, G.~A. Mena~Marug\'an, and J.~Olmedo,
\newblock {\em Hybrid quantization of an inflationary model: The flat case},
\newblock Phys. Rev. D {\bf 88}, 044013 (2013), arXiv:1307.5222.

\bibitem{Gomar:2014faa}
L.~C. Gomar, M.~Fern\'andez-M\'endez, G.~A.~M. Marug\'an, and J.~Olmedo,
\newblock {\em Cosmological perturbations in Hybrid Loop Quantum Cosmology:
  Mukhanov-Sasaki variables},
\newblock Phys. Rev. D {\bf 90}, 064015 (2014), arXiv:1407.0998.

\bibitem{Gomar:2015oea}
L.~C. Gomar, M.~Mart\'\i{}n-Benito, and G.~A.~M. Marug\'an,
\newblock {\em Gauge-Invariant Perturbations in Hybrid Quantum Cosmology},
\newblock JCAP {\bf 06}, 045 (2015), arXiv:1503.03907.

\bibitem{Martinez:2016hmn}
F.~B. Mart\'\i{}nez and J.~Olmedo,
\newblock {\em Primordial tensor modes of the early Universe},
\newblock Phys. Rev. D {\bf 93}, 124008 (2016), arXiv:1605.04293.

\bibitem{ElizagaNavascues:2020uyf}
B.~Elizaga~Navascu\'es and G.~A.~M. Marug\'an,
\newblock {\em Hybrid Loop Quantum Cosmology: An Overview},
\newblock Front. Astron. Space Sci. {\bf 8}, 81 (2021), arXiv:2011.04559.

\bibitem{Bojowald:2008gz}
M.~Bojowald, G.~M. Hossain, M.~Kagan, and S.~Shankaranarayanan,
\newblock {\em Anomaly freedom in perturbative loop quantum gravity},
\newblock Phys. Rev. D {\bf 78}, 063547 (2008), arXiv:0806.3929.

\bibitem{Cailleteau:2012fy}
T.~Cailleteau, A.~Barrau, J.~Grain, and F.~Vidotto,
\newblock {\em Consistency of holonomy-corrected scalar, vector and tensor
  perturbations in Loop Quantum Cosmology},
\newblock Phys. Rev. D {\bf 86}, 087301 (2012), arXiv:1206.6736.

\bibitem{Cailleteau:2011kr}
T.~Cailleteau, J.~Mielczarek, A.~Barrau, and J.~Grain,
\newblock {\em Anomaly-free scalar perturbations with holonomy corrections in
  loop quantum cosmology},
\newblock Class. Quant. Grav. {\bf 29}, 095010 (2012), arXiv:1111.3535.

\bibitem{Wilson-Ewing:2015sfx}
E.~Wilson-Ewing,
\newblock {\em Separate universes in loop quantum cosmology: framework and
  applications},
\newblock Int. J. Mod. Phys. D {\bf 25}, 1642002 (2016), arXiv:1512.05743.

\bibitem{Iteanu:2022zha}
S.~Iteanu and G.~A. Mena~Marug\'an,
\newblock {\em{Mass of Cosmological Perturbations in the Hybrid and Dressed Metric
  Formalisms of Loop Quantum Cosmology for the Starobinsky and Exponential
  Potentials}},
\newblock Universe {\bf 8}, 463 (2022), arXiv:2208.01987.

\bibitem{Li:2022evi}
B.-F. Li and P.~Singh,
\newblock {\em Close relationship between the dressed metric and the hybrid
  approach to perturbations in effective loop quantum cosmology},
\newblock Phys. Rev. D {\bf 106}, 086015 (2022), arXiv:2206.12434.

\bibitem{Li:2023res}
B.-F. Li and P.~Singh,
\newblock {\em Alternative effective mass functions in the modified
  Mukhanov-Sasaki equation of loop quantum cosmology},
\newblock Phys. Rev. D {\bf 109}, 066005 (2024), arXiv:2310.18408.

\bibitem{Li:2024xxz}
B.-F. Li, M.~Motaharfar, and P.~Singh,
\newblock {\em{Constraining regularization ambiguities in loop quantum cosmology
  via CMB}},
\newblock Phys. Rev. D {\bf 110}, 066005 (2024), arXiv:2405.12296.

\bibitem{Planck:2018jri}
Planck, Y.~Akrami {\em et~al.},
\newblock {\em Planck 2018 results. X. Constraints on inflation},
\newblock Astron. Astrophys. {\bf 641}, A10 (2020), arXiv:1807.06211.

\bibitem{Planck:2018vyg}
Planck, N.~Aghanim {\em et~al.},
\newblock {Planck 2018 results. VI. Cosmological parameters},
\newblock Astron. Astrophys. {\bf 641}, A6 (2020), arXiv:1807.06209,
\newblock [Erratum: Astron.Astrophys. 652, C4 (2021)].

\end{thebibliography}
\end{document}